\documentclass[conference,letterpaper]{IEEEtran}

\usepackage{amsmath,amsfonts}
\usepackage{algorithmic}
\usepackage{algorithm}
\usepackage{array}
\usepackage{textcomp}
\usepackage{stfloats}
\usepackage{xurl}
\usepackage{graphicx}
\usepackage{cite}
\usepackage{xcolor}
\usepackage{subfigure}

\newcommand{\texorpdfstring}[2]{#1}

\begin{document}

\title{RAPID: Proactive Rate Control for Enabling High Data Rate and Low Latency in 5G Edge Networks}

\author{\IEEEauthorblockN{Hamid Hassani, Sonia M. Heemstra de Groot, Ignas G. Niemegeers, Kishor C. Joshi, George Exarchakos}
	\IEEEauthorblockA{Eindhoven University of Technology, The Netherlands}
        {\tt \{h.hassani,sheemstradegroot,i.g.m.m.niemegeers,k.c.joshi,g.exarchakos\}@tue.nl}
}

\maketitle

\begin{abstract}
% Next-generation applications such as eXtended Reality and interactive media services come with strict requirements in terms of data rate, latency, and reliability.
Many 6G use cases such as tactile Internet and eXtended reality applications would simultaneously require a high data rate and low latency wireless connectivity. 
%While the focus of physical and link layer protocols in 5G has been to create more network capacity, 
% It is increasingly being realized that a wider redesign of transport protocols is required to fulfill these stringent requirements. 
% The existing transport solutions relying on loss and delay are not suitable for applications simultaneously requiring low latency and high data rates. 
It is increasingly being realized that the current transport protocols relying on packet loss and delay are insufficient %not suitable 
to satisfy the data rate and latency requirements of such applications. 
To address this problem, we propose RAPID, a transport layer solution designed for private 5G edge networks, where the bottleneck is located at the base station. We show that by regulating the send rate to maintain a target physical resource block utilization, we can achieve high data rate, low latency communication. We design a simple yet effective policy that adapts to the dynamic nature of wireless channel conditions. 
% Through %extensive NS3 simulations and 
% testbed experiments, we evaluate the performance of our design. % in single- and multi-UE scenarios. 
% For example, in a single-UE scenario, our testbed results demonstrate a significant improvement in one-way delay (around $18$x) when compared to widely used TCP Cubic and BBR, with negligible packet loss and a modest deterioration in goodput (around $20\%$). 
%For example, 
In a multi-UE scenario with two RAPID flows and one TCP Cubic flow, our testbed results show that RAPID flows achieve significant improvements in one-way delay, i.e., $72$x and $55$x lower compared to TCP Cubic, with negligible packet loss at the cost of modest goodput reductions of $16\%$ and $27\%$, respectively.
\end{abstract}

% \begin{IEEEkeywords}
% Transport protocols, send rate control, low latency, 5G edge network.
% \end{IEEEkeywords}

\section{Introduction} \label{sec:intro}
Next-generation applications such as eXtended reality (XR), holographic telepresence, and telemedicine operations come with strict requirements for communication in terms of data rate, latency, and reliability %see, e.g., Table \ref{table:req}, 
\cite{Morin22}. 
%\HH{*For instance, \cite{Saad20} states that to provide a full immersive experience, the data rate should range from $100$Gbps to $1$Tbps, latency should be less than $1$ms, and reliability should achieve a level of $5$ to $7$ nines.} 
% For instance, \cite{Saad20} states that to provide a full immersive experience, the following requirements should be satisfied:
% \begin{table}[hbt]
% \centering
% \caption{QoS Requirements of next-generation XR applications \cite{Saad20}}
% \label{table:req}
% \begin{tabular}{|p{2cm}|p{3.5cm}|}
% \hline
% Data Rate       & $100$Gbps - $1$Tbps       \\ \hline
% Latency         & $< 1$ms                   \\ \hline
% Reliability     & $99.999\%$ - $99.99999\%$ \\ \hline
% \end{tabular}
% \end{table}
These requirements, however, cannot be entirely met by current 5G or 802.11ax (Wi-Fi 6) networks that primarily focus on physical and link layer technologies, see, e.g., \cite{Akyildiz22}. %\cite{Saad20, Akyildiz22, Reshef22}. 
It is increasingly being realized to have well-designed transport layer protocols to fulfill the stringent requirements of next-generation applications. 
For example, \cite{Xie22} has compared the end-to-end performance of multiple congestion control (CC) algorithms %using the Pantheon platform, see \cite{Yan18}, 
in a commercial LTE network (AT\&T), and showed that the performance variance among the CC mechanisms might be significant. This measurement, along with many similar measurements conducted using 3GPP or IEEE 802.11 access technologies, see, e.g., \cite{ExLL18,Hassani21}, %,PBE20,ExLL18}, %,Kim14},
provide additional evidence %regarding the insufficiency of relying exclusively on physical and link layer solutions.
% that physical and link layer solutions are necessary, yet insufficient.
that highlights the limitations of existing transport layer solutions in satisfying the demanding requirements of next-generation applications.

The requirement for major upgrades to current transport protocols is also reflected in initiatives such as L4S \cite{L4SNokia}, Google BBR \cite{BBR}, Google QUIC \cite{RFC9000}, and the OpenFastPath Alliance \cite{OFP}. 
Furthermore, ETSI has %recently 
set up the Industry Specification Group on Next Generation Protocols (NGP ISG) to study the next-generation protocols for 5G \cite{NGP}.

According to the applied congestion signal, %\footnote{An indicator used by congestion control mechanisms to detect congestion within the network.}, 
the CC mechanisms at the transport layer are categorized into two main groups: loss-based and delay-based approaches, see Section \ref{sec:relatedwork}. 
%\HH{*For loss-based solutions such as TCP Cubic~\cite{Ha08} and TCP BIC~\cite{Xu04}, packet loss serves as the congestion signal. 
%On the other hand, delay-based solutions like TCP BBR~\cite{BBR} and TCP NewReno \cite{RFC6582}, rely on delay measurements, e.g., round-trip time (RTT), to infer the queue buildup at the path bottleneck.} 
% Another notable approach for congestion control in cellular networks is NG-Scope~\cite{Xie22} and , which employs several passive sniffers located in different sites to measure cell-wide resource utilization and gauge available capacity. 
These solutions, however, %are not suitable \HH{or ``may prove insufficient''?} 
may prove insufficient for next-generation applications that would require both high data rate and low latency, see, e.g., \cite{Vleeschauwer24}. The use of packet loss as a congestion signal is not ideal for achieving low latency and high data rate, as it triggers action, i.e., adjusting the send rate, only after queue overflow occurs, leading to high queueing delay (QD). % \cite{Tanenbaum21}. 
%\HH{*Also, the transport layer solutions which rely on packet loss as the congestion signal cannot differentiate if the loss is due to poor channel quality or queue overflow.} 
On the other hand, for delay-based approaches, measurements of one-way delay (OWD) might be used to infer the onset of queueing and adjust the send rate, but measuring OWD is known to be challenging due to, e.g., clock offset and skew between server and client \cite{Hassani21}, as is inference of queueing from OWD. The use of round-trip time (RTT) to estimate the onset of queueing is also known to be inaccurate when there is queueing in the reverse path, i.e., asymmetric forward and reverse paths \cite{Chefrour21}. 
% Alternative approaches, such as using network-wide sniffing~\cite{PBE20,Xie22} or using explicit congestion notification (ECN) protocols like L4S~\cite{L4SNokia}, are impractical due to deployment and upgrade difficulties.

There are alternative approaches, such as using network-wide sniffing \cite{Xie22,Wan24b} %\cite{PBE20,Xie22,Wan24b} 
to measure cell-wide resource utilization and gauge available capacity, or using explicit congestion notification (ECN) protocols. % like L4S~\cite{L4SNokia}. 
However, these strategies are often impractical due to deployment and upgrade difficulties. 
For example, solutions that rely on ECN, such as L4S \cite{L4SNokia}, require extensive updates to network infrastructure and end-user devices, making widespread adoption challenging.

To address the above challenges, we propose RAPID, a novel user-space transport solution designed to achieve high data rate and low latency communication over 5G edge networks. %, where the path bottleneck is the radio link control (RLC) buffer at the BS. 
% \HH{In such networks, computing resources and data storage are deployed close to the elements of the network edge, such as base stations (BSs) \cite{Pham20}, to optimize the performance for ultra-low latency and high data rate services \cite{ETSI_18}.} 
Our main contributions can be summarized as follows:
\begin{itemize}
    \item We show that by regulating the downlink send rate to maintain a target physical resource block (PRB) utilization, it is possible to avoid persistent queue backlogs at the base station (BS) and, hence, to realize high data rate, low latency communication in a robust and practical fashion.%to realize high data rate and low latency data transmission in a robust and practical fashion.
    
    \item % We show that the proposed solution does not introduce any extra overhead. 
    This approach offers significant advantages in terms of 
    %Furthermore, another great advantage of this approach is its 
    flexibility, support for innovation, and ease of rollout, due to the high efficiency achievable in user space, supported by modern kernels. 
    % \HH{This has been significantly facilitated by the high efficiency achievable in user space, supported by modern kernels.}
    It can be seamlessly integrated as an application on end devices, i.e., servers and user equipment (UEs), without requiring any changes to the existing BSs.
    
    % \item \HH{remove?} We present a prototype transport layer implementation of our low latency rate allocation approach and evaluate its performance using both NS3 simulations and 5G testbed experiments under real radio channel conditions.
    % in both the NS3 network simulator and a 5G testbed and use this to evaluate performance under real radio channel conditions.
    
    % \item Through NS3 simulations in a single-UE scenario, we demonstrate a significant improvement in OWD (around $60$x) when compared to TCP Cubic and BBR, with negligible packet loss ($0.01\%$) and only a slight reduction in goodput ($3\%$).

    \item 
    % The 5G testbed evaluation shows an $18$x reduction in OWD compared to TCP Cubic and a $44$x reduction compared to BBR in single-UE scenarios, with around $20\%$ reduction in goodput compared to both mechanisms. 
    The 5G testbed evaluation using OpenAirInterface (OAI) and Ettus USRPs, see Fig. \ref{fig:SysModv2} (b), in a multi-UE scenario, with two UEs operating RAPID and one UE using a TCP Cubic flow, shows that %under varying channel conditions, 
    the RAPID flows achieve $72$x and $55$x reductions in OWD with negligible packet loss, while incurring $16\%$ and $27\%$ deteriorations in goodput, respectively, compared to the TCP Cubic flow.
    
\end{itemize}

Note that the 5G QoS model also supports non-guaranteed bit rate (Non-GBR) flows for low-latency eMBB applications such as XR \cite{ETSI_23.501}. For instance, Non-GBR QoS flows with a 5G QoS Identifier (5QI) value of $80$ are proposed for XR applications with a packet delay budget (PDB) of $10$ms %\footnote{While Non-GBR flows can experience congestion-related packet drops and delays, in uncongested network conditions, $98\%$ of packets should not experience a delay exceeding the 5QI's PDB \cite{ETSI_23.501}.} 
when the network is not congested. 
While these QoS mechanisms provide performance targets, they alone cannot prevent network congestion. 
Our transport layer solution complements the 5G QoS model by enabling end devices to proactively adjust their send rates to avoid congestion and maintain optimal performance, % \cite{Tanenbaum21}, 
facilitating the realization of the performance targets specified by 5QI parameters. 

The remainder of this paper is structured as follows: 
In Section \ref{sec:relatedwork}, we discuss related work. 
% We recap some essential preliminaries in Section \ref{sec:prelim}. 
Next, we present the system architecture in Section \ref{sec:system}, followed by an analysis of latency in Section \ref{sec:delay}. We then explain how PRB utilization can be used to infer the onset of queueing at the BS in Section \ref{sec:prb}. In Section \ref{sec:ctrl}, we design a control policy to achieve high data rate and low delay communication in private 5G edge networks. 
%In Section \ref{sec:performance}, we present the NS3 simulation and experimental results for single- and multi-UE scenarios, along with a comparison with TCP Cubic and BBR. 
In Section \ref{sec:performance}, we present experimental results. %for a multi-UE scenario. %, along with a comparison with TCP Cubic and BBR. 
Next, we discuss limitations and future work in Section \ref{sec:discussion} and conclude the paper in Section \ref{sec:conclusion}.

%\vspace{-10pt}

\section{Related Work} \label{sec:relatedwork}

\subsection{Congestion Control}
%\HH{*The only effective way to manage congestion is for the transport protocols to reduce the rate at which packets are sent into the network \cite{Tanenbaum21}.} 
The delay-based CC mechanisms such as % \HH{wrong} TCP NewReno \cite{RFC6582} 
TCP Vegas \cite{vegas} and TCP BBR \cite{BBR} use RTT (/minimum RTT) measurements to adjust the send rate and avoid the queue build-up in the path bottleneck links. This way, they target low latency communication; however, several studies in cellular networks have shown that they fall short in achieving good performance, see, e.g., \cite{claw17, ExLL18, PBE20, Xie22}, because, e.g., the forward path and reverse path are not typically identical due to queueing in the reverse path \cite{Chefrour21}. %\cite{Pathak08}.

The CC mechanisms that use packet loss as the congestion signal, e.g., TCP BIC \cite{Xu04} and TCP Cubic \cite{Ha08}, are not suitable for low delay communication since they increase the send rate to fill up the bottleneck queues until they see packet losses, which result in high latency.
 
%\HH{*The ECN protocols, such as L4S \cite{L4SNokia}, require changes in the existing network equipment. However, introducing modifications to the existing %physical and 
%link layer solutions requires a widespread software upgrade across all network nodes, such as APs and BSs. This process is notably expensive and resource-intensive, which can ultimately restrict the agile integration of innovative solutions.}

%\HH{*While these transport layer solutions are typically executed in kernel space, due to the high efficiency achievable in user space and the support of modern kernels, there is a growing interest in user space transport solutions \cite{Hassani21}. This interest is due to, e.g., their flexibility and ease of rollout. A well-known example of these user space solutions is Google QUIC \cite{RFC9000}.}

TCP BBR \cite{BBR} uses the measurements of RTT and throughput to estimate the bottleneck bandwidth and avoid queue build-up. While the Linux kernel only supports BBRv1, Google plans to submit BBRv3 for inclusion in mainline Linux \cite{BBRv3}. Note that BBR is not specifically designed for wireless networks, and multiple studies, see, e.g., \cite{Im19, Hassani21, PBE20, Xie22}, show that achieving low latency in 3GPP and IEEE 802.11 networks is challenging via BBR. Once Linux (and network simulators like NS3) upgrades to BBRv3, the performance of the latest release of this rate control mechanism can be widely evaluated; however, initial work like \cite{Zeynali24} raises concern about its performance.

\subsection{Congestion Control for LTE/5G Networks}
The primary focus of relevant literature is on various approaches to estimate the available bandwidth to, e.g., adjust the TCP congestion window (CWND). 
For example, piStream \cite{piStream15} adopts an energy-based spectrum monitor to estimate the available bandwidth in LTE networks. While it does not require decoding the control channel, such power measurements are vulnerable to interference and thus result in significant estimation error \cite{Xie22}. In addition, this method needs fine-grained per-subcarrier channel information, which can be acquired only via radio monitoring software \cite{claw17}.

\cite{claw17} presents a tool called CLAW that can estimate the LTE cell-wide resource utilization based on the physical layer information, such as RSRQ (Reference Signal Received Quality), available on unmodified commodity (but rooted) mobile handsets. CLAW needs around one RTT to estimate the available bandwidth. 
However, results in \cite{claw17} show that the predictive power of signal strength is relatively limited \cite{PBE20}. Furthermore, energy measurements have typically resulted in significant estimation errors due to interference \cite{Xie22}.

ExLL \cite{ExLL18} analyzes the patterns in packet reception in short time windows, i.e., at an LTE subframe granularity, to infer the LTE link bandwidth. This transport layer approach relies on end-to-end statistics so that it is sensitive to network dynamics and can potentially estimate the available bandwidth inaccurately \cite{PBE20}.

A state-of-the-art solution, as outlined in \cite{PBE20,Xie22} for LTE and \cite{Wan24a,Wan24b} for 5G, proposes to increase the send rate based on the unallocated PRBs in the previous slots to maximize throughput and maintain low delay while it does not explicitly discuss its rate control policy. This approach necessitates knowledge of cell-wide PRB utilization, which they obtained by introducing passive sniffers in different locations to capture and decode the control channel over the air interface \cite{PBE20,Xie22,Wan24a,Wan24b}. Nonetheless, this approach has performance concerns, and decoding the control channels is not a practical solution \cite{piStream15}. % , it has only been tested for LTE networks.

\section{System Architecture} \label{sec:system}
\begin{figure}[tbp]
\centering
\hspace{-25pt}
\subfigure[System model]{
\includegraphics[width=0.55\columnwidth]{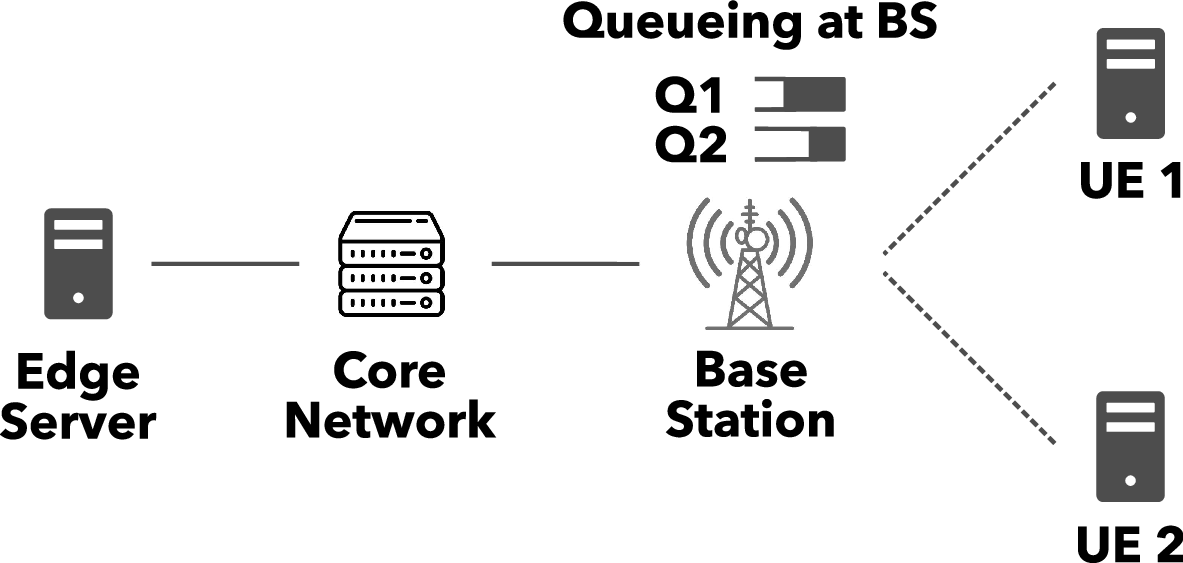}
}
% \hspace{0.01\columnwidth}
\hspace{-5pt}
\subfigure[5G testbed]{
\includegraphics[width=0.38\columnwidth]{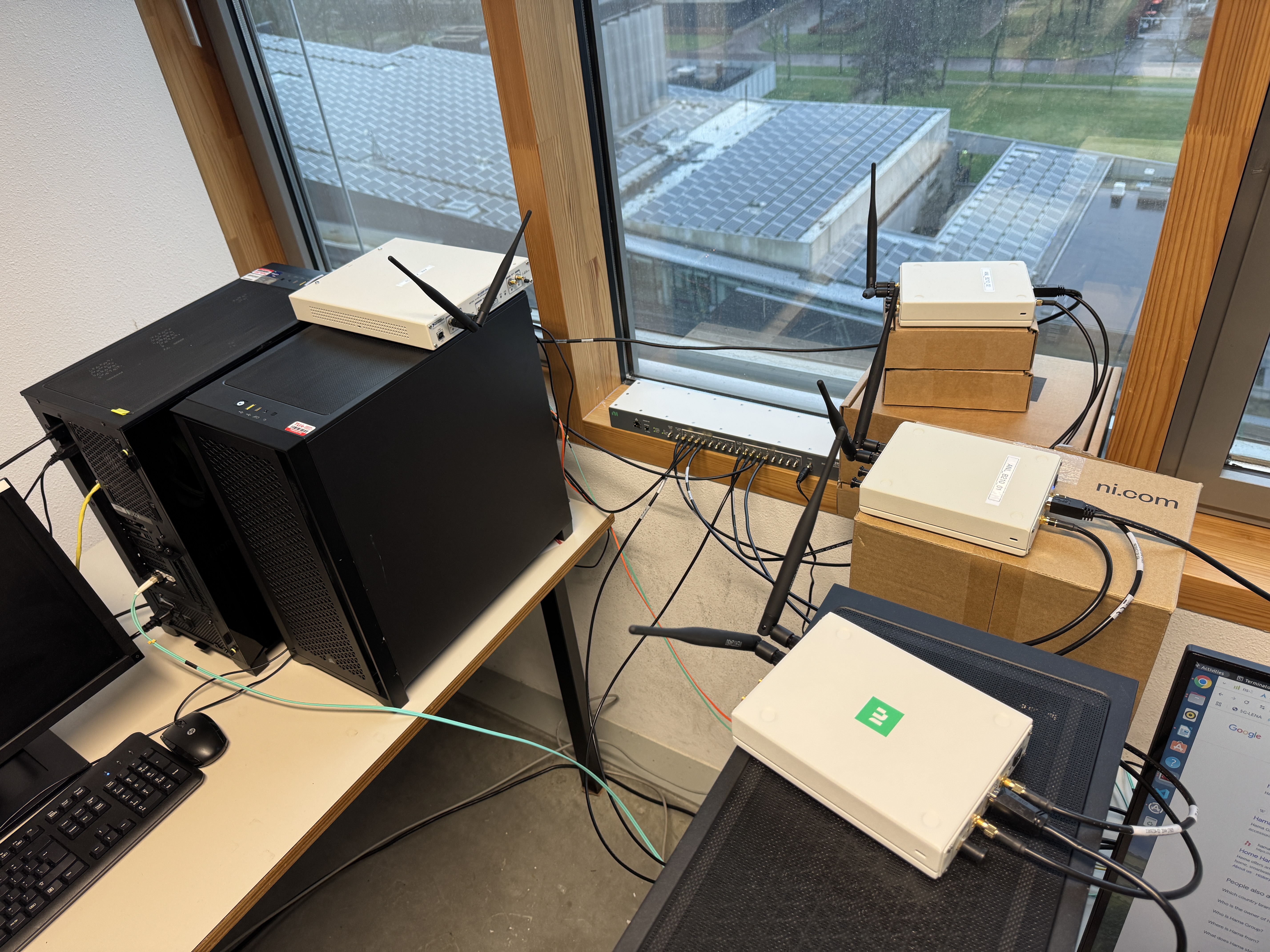}
}
\hspace{-25pt}
\caption{(a) Edge transport architecture with a bottleneck in the wireless link, i.e., at the RLC layer of the BS, as indicated on the schematic. (b) 5G testbed.}
\label{fig:SysModv2}
\end{figure}

% In this paper, we consider the edge architecture of the type illustrated in Fig. \ref{fig:SysModv2} (a), where the BS serves as the bottleneck in an end-to-end path from the edge server (ES) to UEs. This is due to the rate mismatch between the wired backhaul and the air interface. 
In this paper, we consider the edge architecture of the type illustrated in Fig. \ref{fig:SysModv2} (a), where the edge server (ES) and the user plane function (UPF) are collocated with the BS, see, e.g., Fig. 3 in \cite{ETSI_18}. 
We assume that the BS serves as the bottleneck in an end-to-end path from the ES to UEs, due to the rate mismatch between the wired backhaul and the air interface. 
% The locations of the data networks and the UPF are a choice of the network operator and the network operator may choose to place the physical computing resources based on technical and business parameters such as available site facilities, supported applications and their requirements, measured or estimated user load etc \cite{ETSI_18}. 
As depicted in Fig. \ref{fig:SysModv2} (a), the BS adheres to the per-UE queueing policy, meaning that each UE has its dedicated queue at the BS \cite{Irazabal24}. 
% This motivates us to focus on client-side solutions.

Our interest is in achieving high data rate and low latency in 5G edge networks by effectively managing end-to-end connections via an enhanced transport layer protocol. 
The key concept involves preventing the buildup of a persistent queue backlog at the BS in the RLC buffer to keep the %end-to-end delay\footnote{The one-way delay between the edge application and the UE.} low. \HH{or just ``to keep the OWD low?''}
OWD between the edge application and the UE low while fully utilizing the available link capacity.

% \begin{figure}[tbp]
% \centering
% % \hspace{-25pt}
% \includegraphics[width=0.75\linewidth]{SysModv1.eps}
% \caption{Edge transport architecture with a bottleneck in the wireless link, i.e., at the RLC layer of the BS, as indicated on the schematic.} %(\HH{Hamid: mirror this figure to align with Fig \ref{fig:SysModv1}})}
% \label{fig:SysModv2}
% \end{figure}

% \begin{figure}[tbp]
% \centering
% \hspace{-25pt}
% \subfigure[System model]{
% \includegraphics[width=0.60\columnwidth]{SysModv1.eps}
% }
% % \hspace{0.01\columnwidth}
% \hspace{-5pt}
% \subfigure[User-plane protocol stack]{
% \includegraphics[width=0.32\columnwidth]{TBv2.png}
% }
% \hspace{-25pt}
% \caption{(a) Edge transport architecture with a bottleneck in the wireless link, i.e., at the RLC layer of the BS, as indicated on the schematic. (b) Abstract resource allocation schematic at a 5G BS.}
% \label{fig:SysModv2}
% \end{figure}

% While our current focus is on the downlink, extending the results to the uplink is relatively a straightforward process. The main difference is that, in the case of the uplink, each UE has access only to its own link layer information, such as PRB utilization. Hence, to determine the sending rates of UEs under the same administration, the information from all UEs needs to be sent to the edge server as feedback (via the uplink). The edge server can then adjust the sending rates and inform the UEs to take the necessary actions.

% While our current focus is on the downlink, extending this approach to the uplink is relatively straightforward. The key difference is that each UE has access only to its own link layer information, such as PRB utilization. This information needs to be sent as feedback to the ES, which then adjusts and distributes updated uplink send rates to each UE for the subsequent time interval.

While our current focus is on the downlink, extending this approach to the uplink is a relatively straightforward process. The key difference is that for the uplink, the ES first updates the uplink send rates of UEs and then distributes this information to UEs for the subsequent time interval.

\subsection{Assumptions} \label{sec:assumptions}

\subsubsection{System Model}
The network under consideration is a private 5G network where all traffic flows are under the administration of the ES. The ES is located close to the end users, i.e., at the network edge, and has a centralized overview of network flows. %under its administration. 
The ES can also function as a proxy gateway, routing traffic flows to and from UEs. This enables the implementation of novel transport layer protocols over the path between the proxy and UEs. 
%\HH{should I remove this (repetitive) sentence?} 
% \HH{*The path bottleneck is located in the wireless link.} 
%\HH{*Furthermore, the backhaul network and ES are assumed to be reliable without any packet losses.} 
%In addition, the UEs are considered semi-static, operating without requiring handovers.

% When carrier aggregation in the LTE-Advanced networks is activated, multiple frequency bands (e.g., 2 or 3 bands) can add up and the total bandwidth increases to 20, 30, 40, or 50MHz.

\subsubsection{Data Flow} 
We use the UDP transport protocol to deploy our approach as an application that runs on both the ES and UEs. 
The downlink traffic type is considered to be identical, e.g., best effort, so that all traffic flows towards a UE are carried by a single data radio bearer (DRB), thus sharing the same RLC buffer at the BS. 
In addition, we presume a full-buffer model, where there is always packet(s) queued in the RLC buffer of each UE, precisely at scheduling decisions, see, e.g., \cite{Huang18}. 
% We assume packets are generated periodically, and each packet within a transmission block experiences consistent transmission time from the ES to the BS, transmitted along the same path. 
%\HH{*We consider an ideal backhaul network, and that packets within a TB experience consistent transmission time from the ES to the BS, traveling along the same path.
%% No re-ordering of packets occurs during the transmission from the ES to the base station (BS). 
%Furthermore, all data packets %can
%fit within one maximum transmission unit (MTU) %\footnote{Most commercial networks set their MTU sizes between 1428 and 1500 bytes.}) 
%without requiring segmentation. }
% We also assume that the RLC layer operates in unacknowledged mode, which does not use packet retransmission and is used for real-time services.

\section{Latency Analysis} \label{sec:delay}
To design an effective low latency rate control policy, we focus on the OWD of packets transmitted from the ES to UEs within the 5G edge network presented in Fig. \ref{fig:SysModv2} (a). 
%\HH{Recall that measuring RTT is not an accurate approach to infer the onset of queueing at the BS, see Section \ref{sec:intro}.} 
The OWD includes several random components, including processing time, queueing delay, transmission delay, etc. 
However, %as discussed in Section \ref{sec:prelim}, 
the most significant (and often overlooked) source of delay is typically the QD \cite{L4SNokia} at the RLC layer of the BS \cite{Flavien22,Irazabal24}. Our analysis, therefore, excludes other sources of delay. % Irazabal22
% However, since the wireless link is typically significantly slower than the wired domain, the path bottleneck lies at the entity that holds the last buffer in the wired domain, i.e., the RLC buffer \cite{Irazabal22,Irazabal24}. 
% \HH{*Note that, for ease of presentation, we will refer to this source of delay as QD unless stated otherwise.} 
% \HH{*The queueing delay of a packet is the duration it spends waiting at the RLC queue of the BS until it is delivered to the MAC layer for transmission over the air interface.} 
%This period 
The QD is influenced by various factors, including the underlying access technology and the number of earlier-arriving packets in the same queue awaiting transmission over the wireless link. 
% Recall that the BS follows the per-station queueing policy. 
The QD is also significantly affected by the MAC scheduling policy, the queue occupancy of other users, buffer sizes, and the channel quality of connected UEs to the BS.

\section{Low Delay High Data Rate Operation} \label{sec:prb}
We define the PRB utilization of UE $i$ during the time interval $\tau$, denoted by $\text{PRB}_{i, \tau}$, as follows:
\begin{equation} \label{eq:prb}
\text{PRB}_{i, \tau} = \text{PRB}_{\text{max}, \text{slot}} \cdot \frac{\text{PRB}_{\text{alloc},i, \tau}}{\text{PRB}_{\text{total}, \tau}}
\end{equation}
where $\text{PRB}_{\text{alloc},i, \tau}$ is the number of allocated PRBs to UE $i$ and $\text{PRB}_{\text{total}, \tau}$ represents the total number of PRBs in the time interval $\tau$. 
$\text{PRB}_{\text{max}, \text{slot}}$ denotes the maximum number of available PRBs per slot, which is determined by the bandwidth and the selected numerology.

\begin{figure}[t]
\centering
\hspace{-25pt}
\subfigure[Single-UE Scenario]{
\includegraphics[width=0.49\columnwidth]{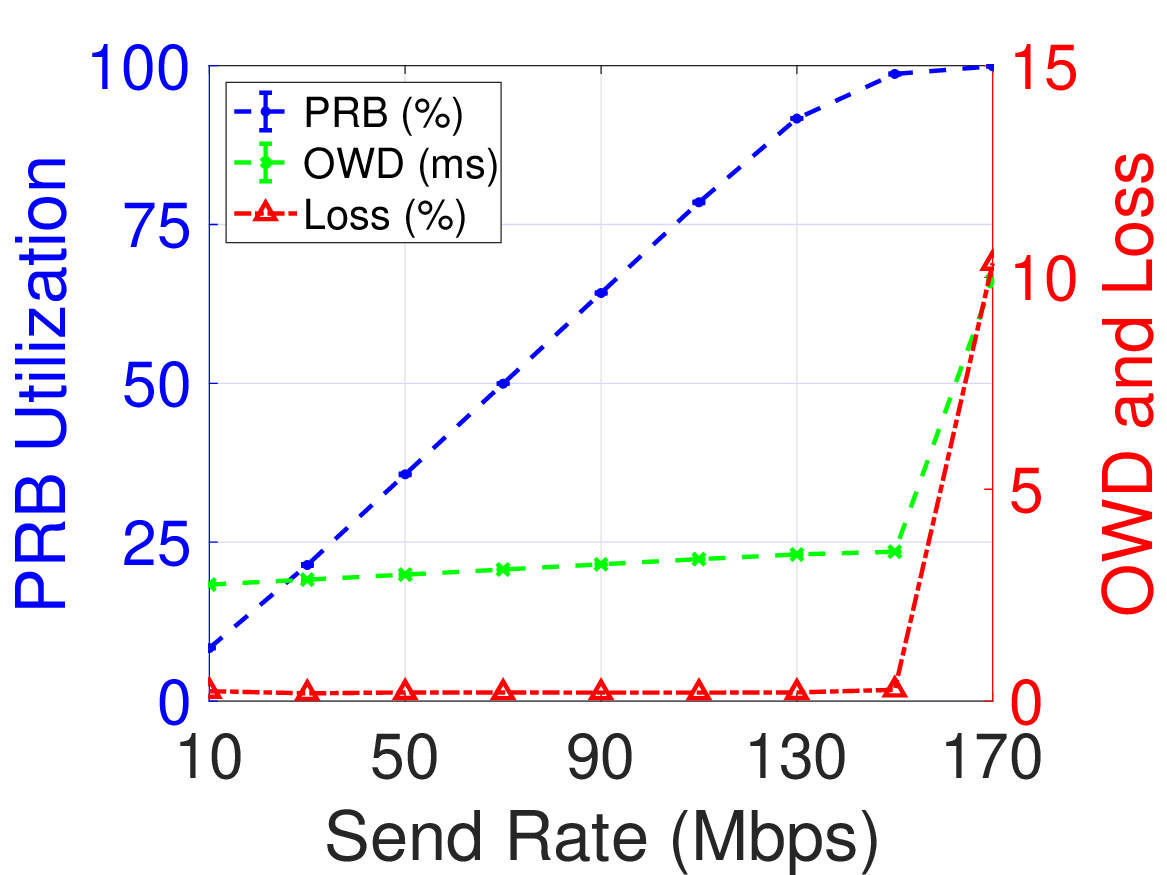}
}
\hspace{-10pt}
\subfigure[Multi-UE Scenario]{
\includegraphics[width=0.49\columnwidth]{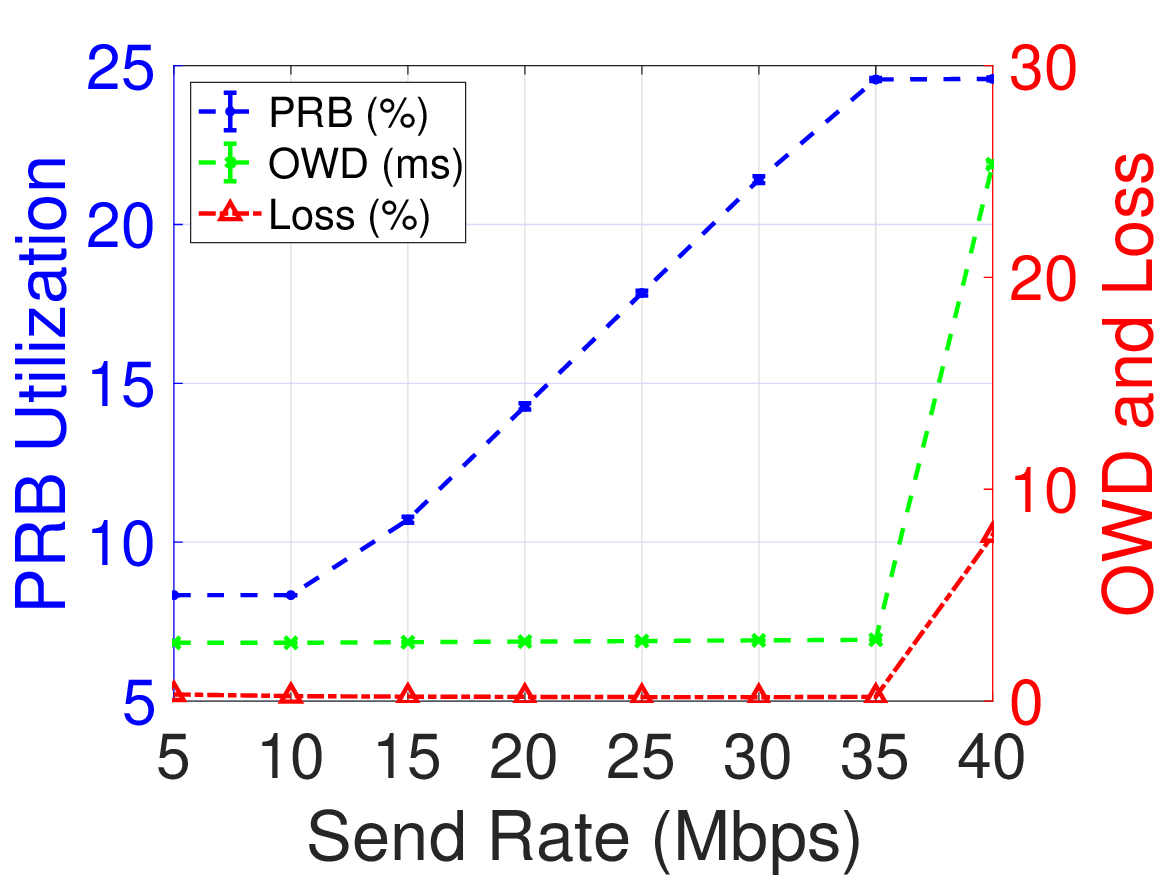}
}
\hspace{-25pt}
% \caption{NS3 simulations using the 5G-LENA module \cite{5gLENA}, unacknowledged mode (RLC) with the default packet scheduler of the 5G-LENA module, i.e., \texttt{NrMacSchedulerTdmaRR}. BW = $40$MHz, UDP traffic, packet size = $1460$B, MTU = $1500$B, n78 band, MCS = $28$, TX power = $20$dB, BS-UE distance $\approx 31$m, $60$sec of simulation. (a) single-UE scenario, (b) $4$ UEs located at the same distance from the BS %at different locations, 
% with equal send rates.}
\caption{NS3 simulations using the 5G-LENA module \cite{5gLENA} with the \texttt{NrMacSchedulerTdmaRR} scheduler. BW $= 40$MHz, UDP traffic, MCS $= 28$, $60$sec of simulation. (a) single-UE scenario, (b) $4$ UEs located at the same distance from the BS %at different locations, 
with equal send rates.}
\label{fig:prb}
\end{figure}

Fig. \ref{fig:prb} depicts the PRB utilization (\%), one-way delay, and packet loss vs. the send rate for a range of network configurations from the NS3 network simulator. A number of key features are evident from the simulation results.

Firstly, as the send rate increases, the PRB utilization increases monotonically until it reaches the maximum value. % supported by the PHY layer. 
Fig. \ref{fig:prb} (a) shows a scenario in which only one UE is connected to the BS, where all available PRBs can potentially be allocated to that UE, resulting in $100\%$ PRB utilization. 
However, in the case of multiple UEs, the PRBs must be shared among UEs via the MAC scheduler. 
Fig. \ref{fig:prb} (b) presents a scenario where identical flows are generated for $4$ UEs, each located at an equal distance from the BS (in a circle) with the same modulation and coding scheme (MCS) and channel model in NS3. 
Note that since the plots for all UEs are nearly identical under these conditions, we only show the plot for one representative UE in Fig. \ref{fig:prb} (b). 
We can observe that Fig. \ref{fig:prb} (a) and (b) follow the same pattern, except that in Fig. \ref{fig:prb} (b), the maximum PRB utilization for each UE reaches $\approx 25\%$. 

Secondly, the OWD increases monotonically with the send rate, initially increasing gradually but then increasing sharply as the send rate approaches the link capacity, leading to the onset of network congestion. Observe that the sharp increase in OWD coincides with PRB utilization approaching its upper limit and with the onset of packet loss. 
We analyze the behavior in Fig. \ref{fig:prb} in more detail. Assume that packets are transmitted by the sender %(/transport layer) 
in a paced fashion. On arriving at the BS, they are queued at the RLC buffer until the next transmission opportunity. The queue occupancy increases roughly linearly since the arriving packets are paced (have roughly constant inter-arrival times). Upon a transmission opportunity, the queued packets are assembled into a transport block (TB) and transmitted by the PHY layer. 
% The TB size in each slot is a random variable, determined by the number of queued packets, the number of allocated PRBs by the MAC scheduler, and the MCS. 
% It, therefore, varies due to, e.g., fluctuating data rate, randomized cross-traffic, and varying channel conditions, 
The maximum number of packets that can be aggregated within a TB is determined by the number of allocated PRBs by the MAC scheduler and the selected MCS. 
If the required number of PRBs by a UE in a given slot is less than the maximum available PRBs, the queue backlog is cleared. Otherwise, packets should wait for the next slot(s) to gain transmission opportunities. 
% The TB size is a random variable due to, e.g., randomized cross-traffic and varying channel conditions, and it depends on the number of allocated PRBs in each slot. 
Thus, queueing delay and PRB utilization are correlated, and this is what we see in Fig. \ref{fig:prb}.

The data in Fig. \ref{fig:prb} suggests that if we could operate the system at a PRB utilization close to the maximum achievable capacity, we can obtain a high send rate while maintaining low delay. 
The approach we suggest here is based on this observation. 
Note that operating at less than the maximum possible PRB utilization incurs a throughput cost and there is therefore a trade-off between delay and data rate. However, Fig. \ref{fig:prb} indicates that this trade-off is quite favorable, i.e., that low delay comes at the cost of only a relatively small reduction in data rate compared to the maximum possible.

\section{Controlling Delay} \label{sec:ctrl}
We proceed by introducing a simple feedback loop that adjusts the downlink send rate at the ES to maintain a selected target PRB utilization. 
This value is determined by, e.g., the QoS requirements of the application. % and the channel quality of the UE. 
% At the UE side, time is partitioned into intervals of duration $\delta$ seconds and we let $\mathcal{T}_{i,j}$ denote the set of TBs transmitted to UE $i$ in the time interval $\delta_{j}$. UE $i$ measures the number of PRBs $\mathcal{P}_{i,f}$ allocated in TB $f$ and reports the average $\mu_{\mathcal{P}_{i}}(\delta_{j}) = \frac{1}{|\mathcal{T}_{i,j}|} \sum_{f \in \mathcal{T}_{i,j}} \mathcal{P}_{i,f}$ back to the edge server every $\delta$ seconds. 

\begin{algorithm}[b]
\begin{algorithmic}
\caption{Feedback loop adjusting send rate $r_i$ to regulate  $\mu_{\text{PRB}_{i}}(\Delta)$.} % Feedback loop
\label{alg:ctrl}
    \STATE Initialize $k \gets 1$
    \WHILE{true}
        \STATE sleep($\Delta$)
        \STATE $\mu_{\text{PRB}_{i}}(\Delta_{k}) \gets \frac{1}{|\mathcal{N}_{i,k}|} \sum_{j \in \mathcal{N}_{i,k}}{\text{PRB}_{i, j}}$
        % \STATE $r_{i} \gets r_i - K \times (\mu_{\text{PRB}_{i}}(\Delta_{k}) - P_{\epsilon,i})$ % \HH{i+1 should be i}
        \STATE $r_{i,k+1} \gets r_{i,k} - K \cdot (\mu_{\text{PRB}_{i}}(\Delta_{k}) - P_{\epsilon,i})$
        \STATE $k \gets k + 1$
    \ENDWHILE
\end{algorithmic}
\end{algorithm}
% for RR the coeeficiient would be 1
% Pe = Pmax * bandwidth efficiency 

At the UE side, time is partitioned into intervals of duration $\delta$ seconds. 
UE $i$ measures its PRB utilization within the interval $\delta_j$ ($\text{PRB}_{i,j}$), see \eqref{eq:prb}, and reports it to the ES via UDP every $\delta$ seconds. 
The ES then uses the received feedback from UE $i$ to adjust its send rate $r_i$. At the ES, time is divided into intervals of duration $\Delta$ seconds ($\Delta > \delta$). 
% Note that while $\Delta = \delta$ reduces resource utilization over the uplink, selecting $\Delta > \delta$ increases reliability in case of packet loss over the uplink. 
We let $\mathcal{N}_{i,k}$ denote the set of feedback reports delivered by 
%received from 
UE $i$ in interval $\Delta_{k}$. 
% Next, we denote $\mu_{\mathcal{P}_{i}}(\Delta_{k}) = \frac{1}{|\mathcal{N}_{i,k}|} \sum_{j \in \mathcal{N}_{i,k}}\mu_{\mathcal{P}_{i}}(\delta_{j})$ the average PRB utilization of UE $i$ in time interval $\Delta_{k}$. 
Next, we denote $\mu_{\text{PRB}_{i}}(\Delta_{k}) = \frac{1}{|\mathcal{N}_{i,k}|} \sum_{j \in \mathcal{N}_{i,k}}{\text{PRB}_{i, j}}$ the average PRB utilization of UE $i$ in the time interval $\Delta_{k}$. 
If the observed $\mu_{\text{PRB}_{i}}(\Delta_{k})$ for UE $i$ is less than its target value $P_{\epsilon,i}$, the ES increases its transmit rate $r_i$ and decreases it if $\mu_{\text{PRB}_{i}}(\Delta_{k}) > P_{\epsilon,i}$. 
This feedback controller is implemented using the pseudo-code shown in Algorithm \ref{alg:ctrl}. 
Note that the feedback gain $K > 0$ balances the convergence rate and stability of the proposed rate adaptation mechanism. 
% \HH{It involves the following design parameters: ES update interval $\Delta$, UE update interval $\delta$, target PRB utilization $P_{\epsilon}$, and feedback gain $K$. %, where 
% Note that $K > 0$ balances the convergence rate and stability of the proposed rate adaptation.}

% In this paper, we assume that UDP traffic flows under RAPID have identical QoS requirements and, therefore, share the same target PRB utilization $P_{\epsilon}$. 
% However, in scenarios where RAPID flows have different QoS requirements, $P_{\epsilon}$ can be readily adjusted for each UE to optimize the network performance. 
  
% Note that update interval $\delta$ is typically on the order of a few radio frames. %, and since $\delta < \Delta$, we select $\delta$ based on the value of $\Delta$. 
% We consider the choice of these parameters in more detail in Section \ref{sec:conv}, but typical values are $\Delta = 200$ms, $\delta = 50$ms, $K = 0.5$, and $P_{\epsilon} = 0.9 \cdot \text{PRB}_{\text{max}, \text{slot}}/ n$ where $n$ is the number of UEs (/traffic flows) connected to the BS. 

% We assume that the value of $n$ is given, i.e., the edge server knows the number of UEs that are connected to (/receiving traffic from) a specific 5G BS. While this assumption is reasonable in private networks, it is not practical in public networks and so designing more sophisticated control strategies is of interest. This is an undertaking in its own right, and we plan to address this in future work. % \HH{airtime fairness!}

\subsection{Selecting Design Parameter \texorpdfstring{$P_{\epsilon}$}{P_epsilon}} \label{sec:pf}
% We consider the choice of $P_{\epsilon} = \alpha \cdot \text{PRB}_{\text{max}, \text{slot}}/ n$, where $\alpha$ is a coefficient that determines the operating point of a traffic flow relative to the congestion point, and $n$ represents the number of traffic flows (/UEs). 

In this paper, without loss of generality, we assume that $m$ RAPID traffic flows have identical QoS requirements and therefore share the same target PRB utilization, i.e., $P_{\epsilon,i} = P_{\epsilon}$ for $i \in \{1, \ldots, m\}$. 
In scenarios where RAPID flows have different QoS requirements, $P_{\epsilon,i}$ can be readily adjusted for each UE accordingly. %to optimize the network performance.} 
To assess the performance of the system near the network saturation point and under full load, we focus on the maximum achievable PRB utilization per UE. 
We set $P_{\epsilon} = \text{PRB}_{\text{max}, \text{slot}} / n$, where $n \geq m$ represents the total number of active downlink flows, including both RAPID and non-RAPID traffic flows. 
% Note that, to assess the performance of the system near the network saturation point and under full load, we are interested in the maximum achievable PRB utilization for each UE. 
Note that, in private 5G networks, the ES is aware of the number of % UEs with active 
traffic flows, %traffic, see Section \ref{sec:assumptions}, and 
thus the value of $n$ is readily available. 
While this may not hold true for public networks, such information can be obtained in public networks through network telemetry tools such as NR-Scope \cite{Wan24a}.

% We consider the choice of $P_{\epsilon} = \text{PRB}_{\text{max}, \text{slot}} / n$, where $n$ represents the number of active UEs (/downlink traffic flows). 
% Note that, to assess the performance of the system near the network saturation point and under full load, we are interested in the maximum achievable PRB utilization for each UE. 

% \HH{The following two paragraphs address two primary concerns raised by the reviewers: (i) the assumption of knowledge of $n$, and (ii) the rationale behind equal PRB allocation among UEs.}

% Note that, in private 5G networks, the ES is aware of the number of UEs with active downlink traffic, see Section \ref{sec:assumptions}, and thus the value of $n$ is readily available. 
% % While this may not hold true for public networks, designing more sophisticated control strategies to address this limitation is of interest. This is an undertaking in its own right, and we plan to address this in future work. 
% % Nevertheless, such information can be obtained in public networks using network telemetry tools, such as NR-Scope \cite{Wan24a}.
% % While this may not hold true for public networks, we plan to design more sophisticated control strategies to address this limitation in future work. 
% % Nevertheless, such information can be obtained in public networks using network telemetry tools, such as NR-Scope \cite{Wan24a}.
% While this may not hold true for public networks, such information can be obtained in public networks through network telemetry tools such as NR-Scope \cite{Wan24a}.

The reason for considering equal PRB allocation among UEs, i.e., dividing $\text{PRB}_{\text{max}, \text{slot}}$ by $n$, is due to the steady-state behavior of widely used proportional-fair (PF) schedulers in cellular networks %\cite{Yin21,Baranasuriya15,piStream15,claw17,Huang18,Raftopoulou21,Flavien22,Kandoi22,Ursu22,Yin23,5gLENA,OAI_PF}. 
\cite{Yin21,claw17}. %,Huang18,Ursu22,OAI}. 
While some proprietary PF schedulers may implement more sophisticated policies, the logarithmic utility function is prevalently applied in the literature for designing PF schedulers \cite{ExLL18}. 
% \HH{in both 3GPP and IEEE 802.11 access technologies to optimize network performance (/throughput) and ensure fairness among clients \cite{Jin04,ExLL18,Huang18,Gringoli21,Raftopoulou21,Kandoi22}.} 
In this paper, we adopt a similar assumption for the MAC scheduler. 
In steady-state conditions, UEs receive equal transmission probabilities, which results in each UE receiving an equal share of PRBs over time under the PF scheduler, see, e.g., \cite{Yin21}. %,Ursu22,OAI}. 
% \HH{I have provided a mathematical proof for this claim in Section \ref{sec:equalprb}.} 
% \HH{*Note that this behavior is similar to that of the round-robin scheduler \cite{Brunello21,Irazabal24}.} 

% This assumption, however, may not precisely reflect all schedulers implemented in proprietary BSs. Thus, our solution may face incompatibility with certain types of schedulers, leading to under- or over-utilization of network resources. Therefore, designing control strategies that can adapt to such discrepancies and be more independent of the MAC scheduler is a significant challenge, and we plan to address this in future work.

% This utility function helps balance maximizing throughput and ensuring fairness among users, making it well-suited for various wireless communication scenarios. Its concave shape naturally promotes proportional fairness by giving diminishing returns as more resources are allocated, encouraging resources to be distributed more fairly, especially to users with lower throughput.

% Since UEs typically experience different channel conditions, the scheduler shall attempt to exploit the temporal variations by allocating resources to users with the best conditions for the upcoming time period \cite{Yin21}. 

\subsection{Selecting Design Parameters \texorpdfstring{$\Delta$}{Delta} and \texorpdfstring{$K$}{K}} \label{sec:conv}

% \subsection{NS3 Implementation}
We implement our approach in NS3 (v3.41), %the NS3 (v3.41) network simulator, 
using the 5G-LENA module \cite{Patriciello19}. % for the simulation of 3GPP NR non-standalone 5G networks. 
% It periodically adjusts the send rate of \texttt{udp-client} applications at the ES connected to the core network, i.e., PGW, based on the received feedback. 
% Each UE receives a UDP traffic flow at a \texttt{udp-server} application that we modified to collect PRB utilization statistics and periodically transmit this information to the edge server at intervals of $\delta$ms. 
% The feedback controller then adjusts the send rate of \texttt{udp-client} applications at the ES connected to the core network, i.e., PGW, based on the received feedback according to Algorithm \ref{alg:ctrl} at intervals of $\Delta$ms. 
Each UE runs a modified \texttt{udp-server} application, which collects PRB utilization statistics and periodically sends this information to the ES every $\delta$ seconds. The feedback controller at the ES adjusts the send rates of \texttt{udp-client} applications according to Algorithm \ref{alg:ctrl} every $\Delta$ seconds, based on the feedback received from UEs.

% For the propagation loss model, we use the \texttt{InH\_OfficeOpen\_Los} model for a channel with line-of-sight characteristics operating over $40$MHz bandwidth in the n78 frequency band. The numerology is set to $0$, i.e., $15$kHz subcarrier spacing similar to LTE, in the following simulations unless stated otherwise. We use the default packet scheduler of the 5G-LENA module, \texttt{NrMacSchedulerTdmaRR}. Note that the 5G-LENA implementation follows a per-UE queueing policy at the BS.

%\HH{*We use a channel with line-of-sight characteristics operating over $40$MHz bandwidth in the n78 frequency band. The numerology is set to $0$, i.e., $15$kHz subcarrier spacing (SCS) similar to LTE. We use the default packet scheduler of the 5G-LENA module, \texttt{NrMacSchedulerTdmaRR}. 
%% Note that the 5G-LENA implementation follows a per-UE queueing policy at the BS.
%Furthermore, we configure the traffic type as eMBB and enable unacknowledged mode (UM) for the RLC layer.}

\begin{figure}[tb]
\centering
\hspace{-25pt}
\subfigure[Impact of $K$ ($\Delta = 200$ms)]{
\includegraphics[width=0.53\columnwidth]{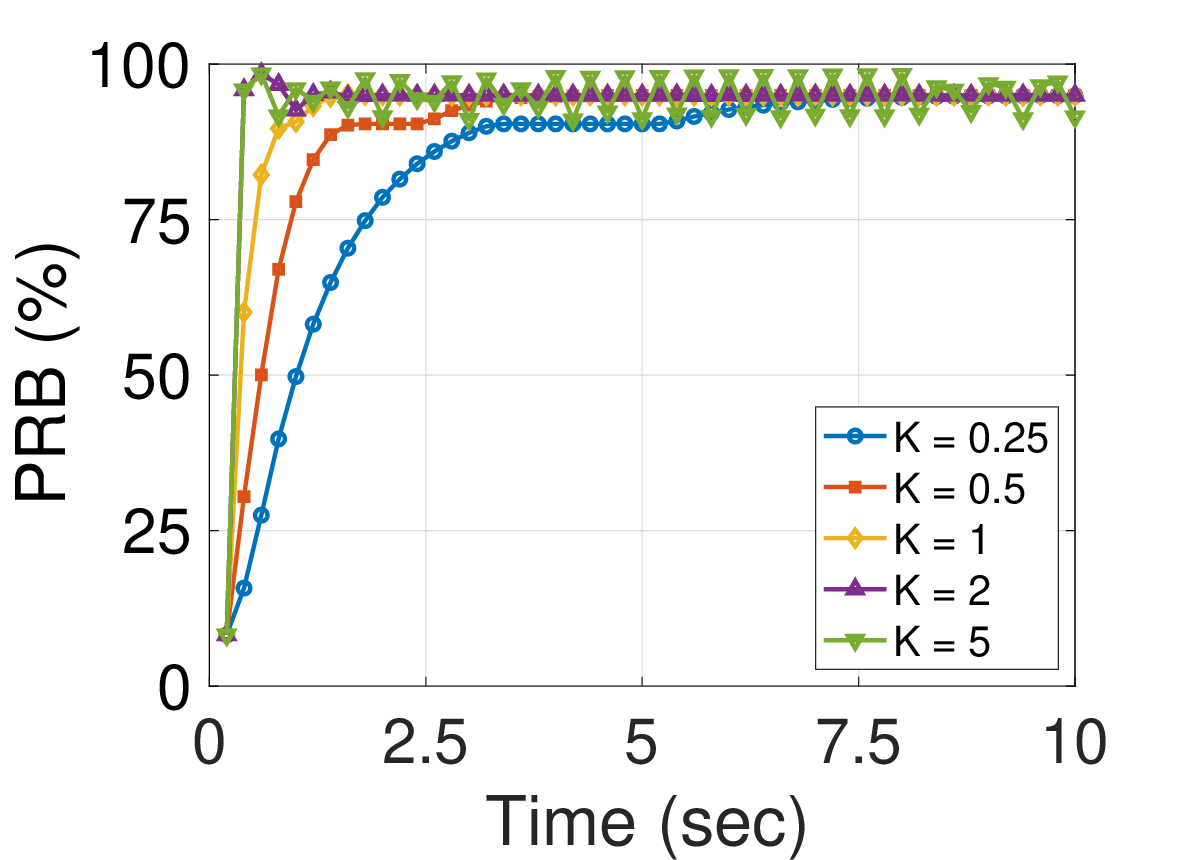}
}
\hspace{-20pt}
\subfigure[Impact of $\Delta$ ($K = 1$)]{
\includegraphics[width=0.53\columnwidth]{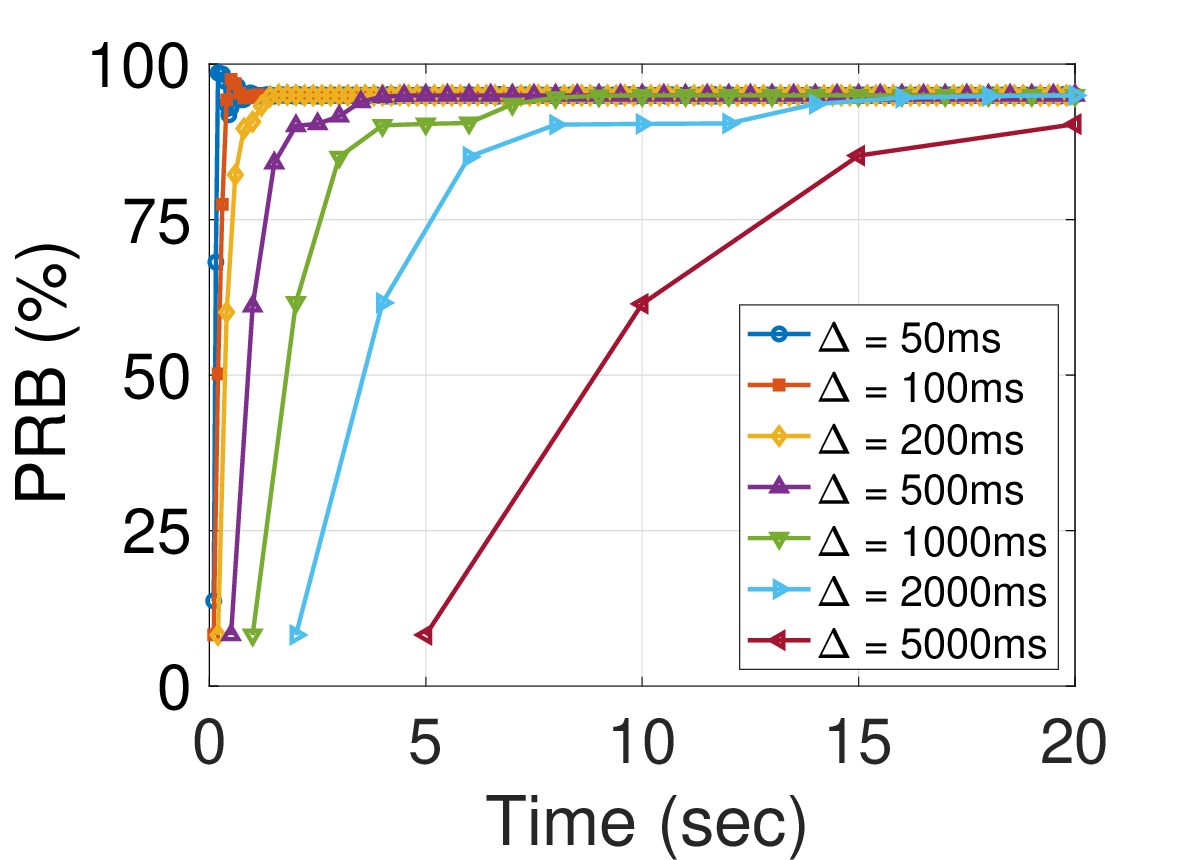}
}
\hspace{-25pt}
% \caption{Convergence rate vs. feedback gain $K$ and update interval $\Delta$. NS3 simulations, single UE, $P_{\epsilon} = 95\%$, $\delta = 50$ms, unacknowledged mode (RLC), BW = $40$MHz, UDP traffic, packet size = $1460$B, MTU = $1500$B, n78 band, MCS = $28$, TX power = $14$dB, BS-UE distance $\approx 31$m. %, $20$sec of simulation.
% }
\caption{Convergence rate vs. feedback gain $K$ and update interval $\Delta$. NS3 simulations, single UE, $P_{\epsilon} = 95\%$, $\delta = 50$ms, BW $= 40$MHz, MCS $= 28$.
}
\label{fig:ctrl}
\end{figure}

% \subsection{Selecting Design Parameters \texorpdfstring{$\Delta$}{Delta} and \texorpdfstring{$K$}{K}} \label{sec:conv}
We expect that the speed at which the PRB utilization and send rate converge to their target values when the ES first starts transmitting to a UE is affected by the choice of feedback gain $K$ and update interval $\Delta$ (with $\delta = 50$ms fixed). 
Fig. \ref{fig:ctrl} shows simulation results demonstrating the transient following startup of a data flow vs. the choice of $K$ and $\Delta$. 
It can be seen from Fig. \ref{fig:ctrl} (a) that as gain $K$ is increased (while $\Delta = 200$ms and $\delta = 50$ms) the time to converge to the target PRB utilization ($P_{\epsilon} = 95\%$) decreases. However, as the gain is increased the feedback loop eventually becomes unstable, e.g., the data for $K = 5$ which shows large, sustained oscillations. %that would obscure the other data on the plot. 
Similarly, it can be seen from Fig. \ref{fig:ctrl} (b) that as the update interval $\Delta$ is decreased the convergence time decreases.

\section{Performance Evaluation} \label{sec:performance}

\subsection{5G Testbed}
Our experimental testbed is a 5G standalone network based on the OpenAirInterface project (branch 2024.w38) for the core, BS, and UEs. 
It is configured to operate in the n78 frequency band using TDD mode with $6$/$3$/$1$ D/U/M configuration, $40$MHz channel bandwidth with $\text{PRB}_{\text{max},\text{slot}} = 106$, %PRBs per slot
single layer transmission, and numerology $1$ ($30$kHz subcarrier spacing with $0.5$ms slot duration). 
This setup adheres to the default configurations of OAI, including the RLC acknowledged mode (AM). 

We use the \texttt{oai-ext-dn} container of OAI as the ES, which runs on the same machine as the 5G core and BS. Each UE operates on a dedicated machine. 
The machines have identical hardware specifications (Intel Core i9-14900K CPUs with $128$GB of DDR5 RAM) and run Ubuntu 22.04.5 LTS. 
The Ethernet NICs are synchronized via the PTP protocol using \texttt{ptp4l} program %\cite{ptp4l} 
over a fiber optic LAN. 
% ptp4l is an implementation of the Precision Time Protocol (PTP) according to IEEE standard 1588 for Linux.
Moreover, we use \texttt{phc2sys} program %\cite{phc2sys} 
to synchronize the system clock to the PTP hardware clock. 

The BS is connected to a USRP X310 software-defined radio (SDR), while the UEs use USRP B210 SDRs to communicate over the air, see Fig. \ref{fig:SysModv2} (b). The USRPs run the open-source UHD driver (v4.6). Furthermore, all USRPs are synchronized using an OctoClock-G CDA-2990 to achieve a high-accuracy time and frequency reference.

We develop the server- and UE-side codes in Python for our prototype implementation of the rate allocation approach presented in Algorithm \ref{alg:ctrl}. 
The ES generates iPerf-like UDP traffic towards the UEs and inserts a sender-side timestamp into the packet payloads. 
Since the machines are tightly synchronized via the PTP protocol, this timestamp can be used to precisely measure the OWD at the UE.

\subsection{Evaluation and Results}
We consider a downlink scenario with three UEs, each assigned an eMBB traffic flow: two UEs receive UDP RAPID flows, i.e., RAPID 1 and RAPID 2, and one UE receives a TCP Cubic flow. % considered as background traffic (/interference).
%\HH{*Note that the RAPID flows have their send rates regulated using the proposed solution in Algorithm \ref{alg:ctrl}.} 
The UEs are placed in different locations, experiencing varying channel conditions. %, see Fig. \ref{fig:3UE} (d).

The key design parameter is the target PRB utilization. 
We set $P_{\epsilon} = 0.9 \times 106 / 3 \approx 32$ PRBs, corresponding to $30\%$ PRB utilization for $n = 3$ UEs. 
The choice of a reduction factor of $0.9$ is intended to enhance system reliability. This decision was based on observations of occasional peaks in OWD associated with high PRB utilization, typically around $95\%$, though such occurrences were rare. 
Through experimental evaluation, we determined that $\Delta = 200$ms, $\delta = 50$ms, and $K = 0.5$ are proper values for our rate adaptation policy in Algorithm \ref{alg:ctrl}. These values are also consistent with the NS3 simulation results, see Fig. \ref{fig:ctrl}.

\begin{figure*}[htp]
  \centering
  \hspace{-26pt}
  \subfigure[Receive rate]{
    \includegraphics[width=0.26\textwidth]{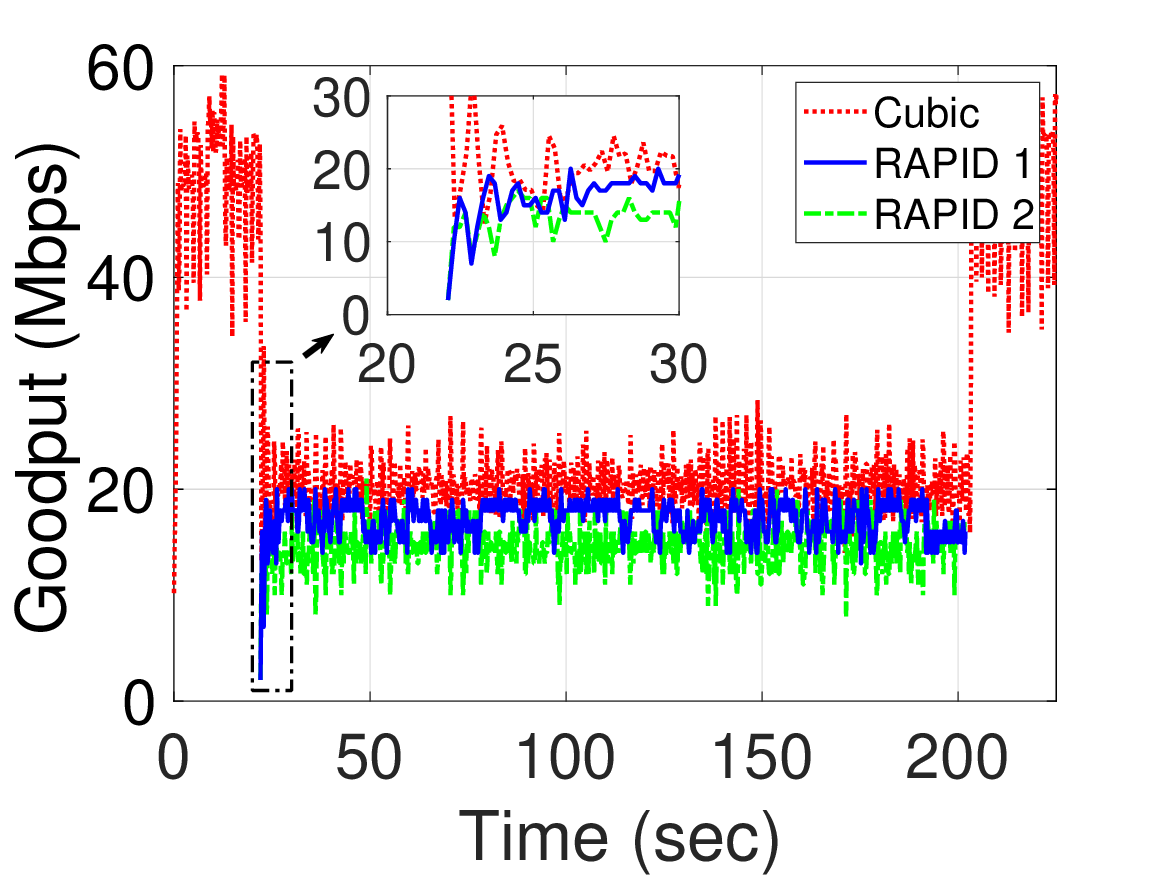}
    }
    \hspace{-20pt}
  \subfigure[One-way delay]{
    \includegraphics[width=0.26\textwidth]{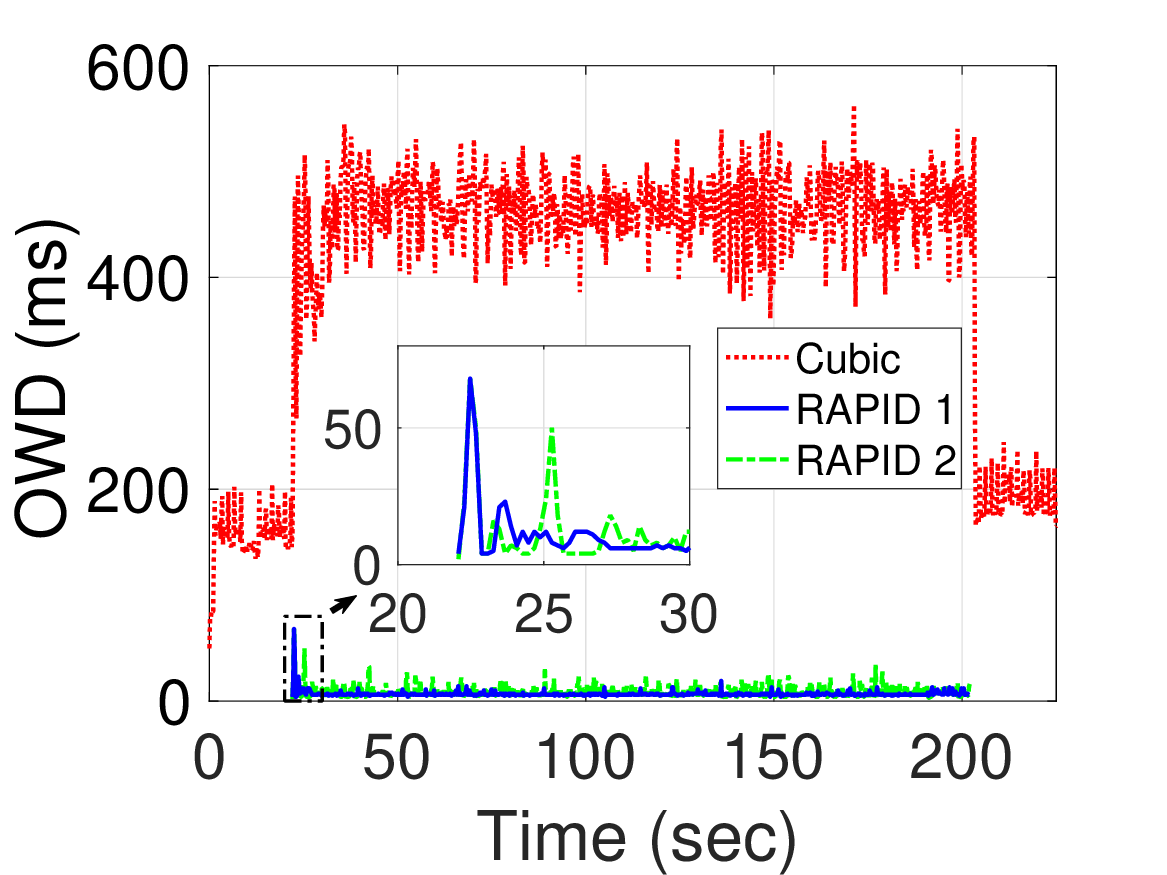}
    }
    \hspace{-20pt}
  \subfigure[PRB utilization]{
    \includegraphics[width=0.26\textwidth]{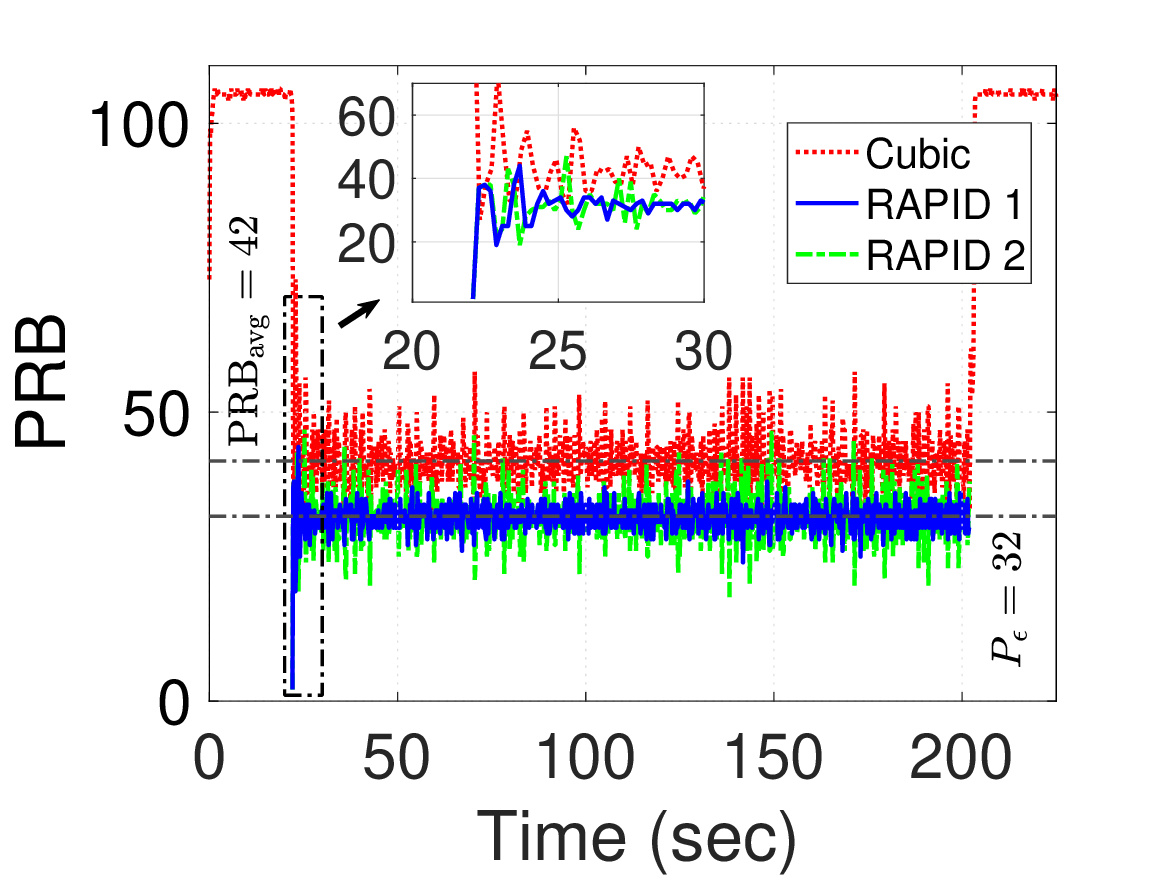}
    }
    \hspace{-20pt}
  \subfigure[DL MCS]{
    \includegraphics[width=0.26\textwidth]{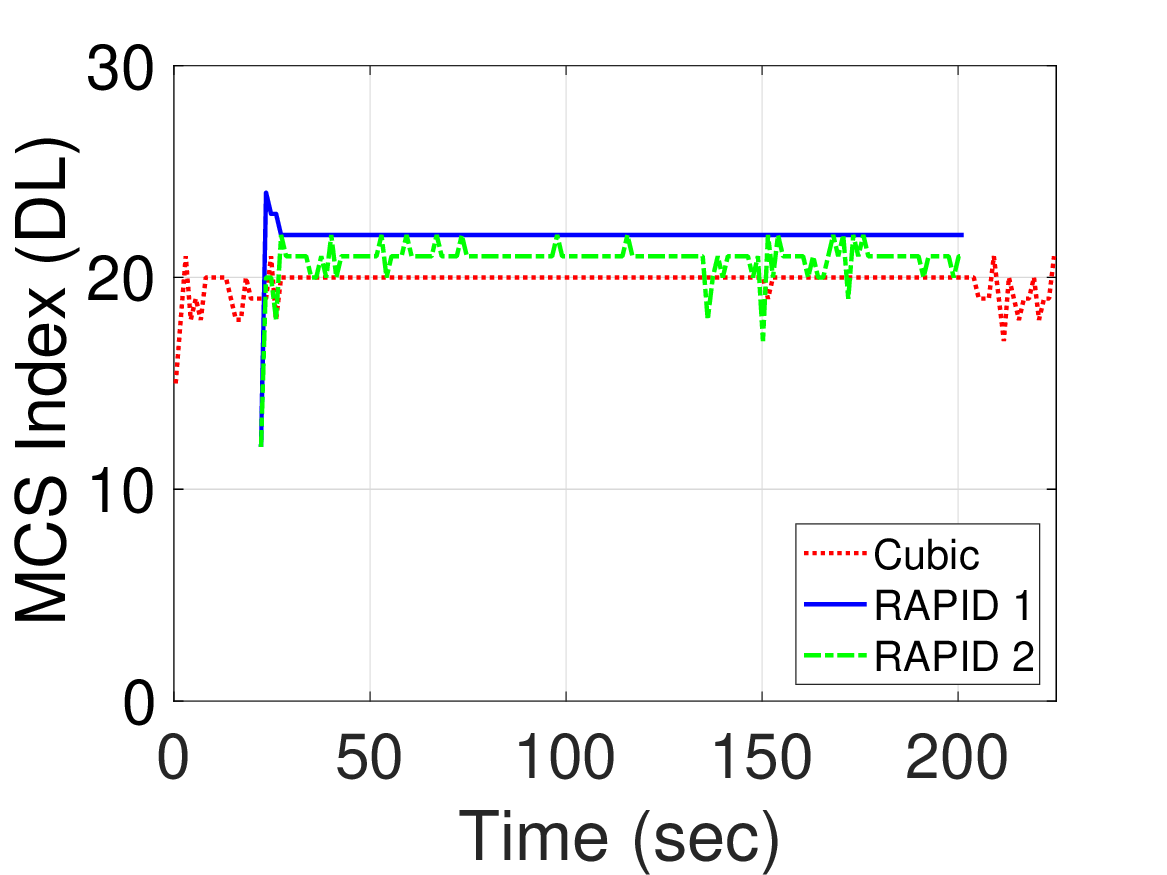}
    }
    \hspace{-26pt}
  \caption{Performance comparison of RAPID with TCP Cubic in a multi-UE scenario. The goodput in (a) is averaged over $200$ms intervals. $\text{PRB}_{\text{max}, \text{slot}} = 106$, $\Delta = 200$ms, $\delta = 50$ms, $K = 0.5$, and $P_{\epsilon} = 30\%$. Experimental results.} 
  %The default RLC buffer size in OAI is $10$MB for the RLC AM mode. }
  \label{fig:3UE}
\end{figure*}

We first initiate the TCP Cubic flow, and after a short interval, the RAPID flows start simultaneously. Fig. \ref{fig:3UE} (a) illustrates the goodput of the traffic flows. 
While the greedy Cubic flow achieves $20.4$Mbps goodput, RAPID 1 and 2 also achieve consistently high goodput around $17.2$Mbps and $14.8$Mbps, respectively, which are $16\%$ and $27\%$ less than the Cubic flow. 
Note that these traffic flows experience different MCS indices over the downlink, see Fig. \ref{fig:3UE} (d). 
Furthermore, the convergence rate of the RAPID flows is fast, occurring within a few seconds after the flows begin, as highlighted in the zoomed inset in Fig. \ref{fig:3UE} (a). 

The reduced receive rate of the RAPID flows (compared with the Cubic flow) can be further analyzed in Fig. \ref{fig:3UE} (c), which depicts the PRB utilization over time. 
The TCP Cubic flow initially utilizes almost all $106$ PRBs when there is no cross-traffic (/the RAPID flows). However, after the initialization of the RAPID flows, the PF scheduler at the BS limits the Cubic flow's resource utilization to an average of $42$ PRBs, while both RAPID flows, with $P_{\epsilon} = 32$, achieve an average of $32$ PRBs, see Fig. \ref{fig:3UE} (c). 
The zoomed inset in Fig. \ref{fig:3UE} (c) showing the $[20, 30]$sec interval illustrates the quick convergence of RAPID flows to $P_{\epsilon}$. 
% \HH{Recall that the lower PRB utilization of the RAPID flows compared to the TCP Cubic is due to the fact that we limit it to $90\%$ of its maximum achievable value in order to keep the OWD low and minimize its peaks, yielding higher reliability in terms of latency.} 

Despite the slightly higher goodput achieved by the TCP Cubic flow, the most significant difference is observed in the OWD measurements in Fig. \ref{fig:3UE} (b). The RAPID flows consistently maintain a much lower OWD compared to the Cubic flow. While the average OWD 
% \HH{*\footnote{We use the \texttt{TCP\_INFO} socket option to %accurately measure the OWD of TCP packets \cite{tcp}.}} %by inserting the sent timestamp of the previously transmitted packet, calculated using the \texttt{tcpi\_last\_data\_sent} field, into the payload of the current packet.} 
of the Cubic flow is $464.9 \pm 33$ms, RAPID 1 and 2 achieve OWDs below $23$ms and $50$ms (after removing the first second of the initialization phase), with averages of $6.5 \pm 2$ms and $8.5 \pm 5$ms, respectively. This represents a $72$x and $55$x reduction in OWD for RAPID 1 and 2, respectively, with significantly smaller fluctuations compared to TCP Cubic. 
The reduced fluctuation in OWD (and goodput) of the RAPID flows, compared to the TCP flow, is particularly important for applications such as XR, video streaming, and online gaming, which require stable performance, see Table 1 in \cite{Gapeyenko23}. 
% \HH{This notable difference in delay characteristics indicates that the proposed control mechanism effectively prevents developing persistent queue backlog at the BS while maintaining competitive receiver rates.} 
Note that since the packet loss is very low, %similar to the single-UE scenario, see Fig. \ref{fig:comp} (c), 
we did not include a packet loss plot.

\section{Discussion} \label{sec:discussion}

% We then compare the performance of RAPID with TCP Cubic \cite{Ha08}, the default CC algorithm used by Linux and Android. In addition, we compare its performance with that of TCP BBR \cite{BBR}, a state-of-the-art CC algorithm developed by Google that also targets high rate and low latency. 
% \HH{should I remove the rest of this paragraph?} The latest version of BBR, i.e., BBRv3, has been introduced recently and is now used for all Google and YouTube public Internet traffic \cite{BBRv3}. However, the mainline NS3 and Linux kernel have not yet been upgraded to support the newer versions of BBR; they currently support only BBRv1.

%\HH{*We compare the performance of RAPID with TCP Cubic, the default CC algorithm in Linux and Android, and Google BBR, a state-of-the-art CC algorithm that also targets high rate and low latency.} 

% \HH{NS3 simulations in a single-UE scenario reveal a significant improvement in OWD (around $60$x) compared to TCP Cubic and BBR, with negligible packet loss ($0.01\%$) and only a slight reduction in goodput ($3\%$). However, due to space constraints, we focused on the more comprehensive and realistic multi-UE scenario.}
 
TCP Prague \cite{Prague} is a recently developed L4S-compatible CC algorithm that also targets low delay, high data rate communication. 
The authors in \cite{Irazabal24} evaluate its performance %the performance of TCP Prague 
in a single-UE scenario using a 5G Open-RAN testbed where they introduce multiple additional components for traffic control. 
These new functionalities may influence the performance of TCP Prague. Furthermore, \cite{Irazabal24} does not provide a detailed evaluation of TCP Prague. 
We are currently working on integrating L4S into our 5G testbed to enable a performance comparison between RAPID and TCP Prague in future work.

% The TCP Prague flow achieves $6\%$ less throughput compared to TCP Cubic with $78\%$ PRB utilization, while the mean delay is reduced by $10.5$x, from $210$ms to $20$ms. 
% Note that drawing a concrete conclusion from these experimental results is challenging since, e.g., the testbed configurations are different and precise definitions of some metrics, such as delay and PRB utilization, are not provided in \cite{Irazabal24}. 
% Nevertheless, without requiring packet marking at the BS as used by TCP Prague, RAPID demonstrates promising performance in comparison. 
% We are currently working on integrating L4S into our 5G testbed to enable a direct performance comparison between RAPID and TCP Prague in future work.
% While testbed configurations are different and precise definitions of some metrics, such as delay and PRB utilization, are not provided in \cite{Irazabal24}, RAPID shows comparable performance without requiring packet marking at the BS.} 
% We are currently working on integrating L4S into our 5G testbed to enable a \HH{direct} performance comparison between RAPID and TCP Prague in future work.

The assumption about the PF scheduler %in 3GPP networks, see Section \ref{sec:pf}, 
in Section \ref{sec:pf} may not precisely reflect all schedulers implemented in proprietary BSs. 
Thus, our solution may face compatibility issues with certain types of schedulers, leading to under- or over-utilization of network resources. Therefore, designing control strategies that can adapt to such discrepancies and be more independent of the MAC scheduler is a significant challenge, and we plan to address this in future work. 
% \HH{Furthermore, we plan to extend our solution to public 5G networks and relax the assumption of having prior knowledge about the number of active traffic flows.} % by designing more sophisticated control strategies. 
% Furthermore, we intend to enable L4S compatibility in our 5G testbed to precisely compare the performance of TCP Prague with RAPID.}

\section{Conclusions} \label{sec:conclusion} % Summary and 
In this paper, we discussed the necessity of well-designed and well-configured transport layer solutions to realize high data rate and low latency communication over 5G edge networks. We showed that the QD at the RLC layer of the BS is correlated with the PRB utilization; therefore, adjusting the send rate to maintain a target PRB utilization can be applied to increase the end-to-end performance, particularly the end-to-end delay. We then proposed a %simple 
rate allocation policy for private 5G edge networks that regulates the downlink send rate based on PRB utilization feedback from UEs. 
% \HH{should I remove this sentance?} 
We implemented the server- and UE-side codes on NS3 %the \HH{NS3} network simulator 
and in our 5G testbed to evaluate the performance of our design. 
%\HH{*The NS3 simulation results for a single-UE scenario demonstrated a $60$x improvement in OWD compared to TCP Cubic and BBR, with negligible packet loss and only a $3\%$ decrease in goodput.} 
The experimental results for a multi-UE scenario, where two RAPID flows coexist with one TCP Cubic flow, showed that RAPID flows achieved $72$x and $55$x lower OWD compared to TCP Cubic, with negligible packet loss, at the cost of modest goodput reductions of $16\%$ and $27\%$, respectively. 
These results further confirmed the feasibility and effectiveness of the proposed approach under real radio conditions. 

\section*{Acknowledgment}
This work was supported by the Distributed Artificial Intelligent Systems (DAIS) project under the Key Digital Technologies Joint Undertaking (KDT JU), Grant Agreement No. 101007273. 
% This work was supported by the DAIS project under the KDT JU grant agreement No. 101007273.

\bibliographystyle{IEEEtran}
\bibliography{references}

\end{document}